\documentclass[floats,aps,unsortedaddress,11pt,tightenlines]{revtex4b1}
\usepackage{cite,latexsym}
\usepackage{graphicx,psfrag}

\begin{document}
\title{Reduction of critical temperatures in pure and thoriated UBe$_{13}$
by columnar defects}

\author{H.A. Radovan}
\author{E. Behne}
\author{R.J. Zieve}
\affiliation{Physics Department, University of California, Davis, CA  95616}
\author{J.S. Kim}
\author{G.R. Stewart}
\affiliation{Department of Physics, University of Florida, Gainesville, FL  32611}
\author{W.-K. Kwok}
\affiliation{Materials Science Division, Argonne National Laboratory,
Argonne, IL  60435}		

\begin{abstract}
We investigate the influence of columnar defects on the superconducting
transition temperatures of UBe$_{13}$ and U$_{0.97}$Th$_{0.03}$Be$_{13}$.  The
defects cause all the transitions to widen and to drop slightly in 
temperature.  Quantitatively, the lower transition in 
U$_{0.97}$Th$_{0.03}$Be$_{13}$ more strongly resembles the single  UBe$_{13}$
transition.
\end{abstract}

\maketitle

Heavy fermion (HF) superconductors are good candidates for  non-$s$-wave
electron pairing.  They display low-temperature power laws in thermodynamic
quantities, unusual behavior of the critical field, and, in the cases of
UPt$_3$ and (U,Th)Be$_{13}$, phase transitions within the superconducting
state.  Impurity scattering, which is sensitive to order parameter symmetry,
has been vigorously investigated in high-temperature superconductors (HTS),
but there has been little corresponding work in HF systems.  Here we report on 
(U,Th)Be$_{13}$ with columnar defects.

Although the most striking effects of columnar defects should appear in their
interactions with vortices, here we focus on their influence on $T_c$.
We compare the behavior of the single transition in pure UBe$_{13}$
with that of both the upper and the lower transitions in 
U$_{0.97}$Th$_{0.03}$Be$_{13}$.   Debate continues on the nature of the lower
thoriated transition, and on which (if either) of the thoriated transitions
can be viewed as an extension of the single pure transition
\cite{Sigrist,Lambert,Zieve,Kromer}. Our present
measurements support identifying the lower thoriated transition with
that of pure UBe$_{13}$.

Our samples are long-term annealed polycrystals, sanded to about 25 $\mu$m 
thickness and 50 $\mu$g.  The defects were created by irradiation with 1.4 GeV
${}^{238}U^{67+}$ ions. Defect densities range from $4.8\times 10^{13}$ to
$2.4\times 10^{15}$ per m$^2$, corresponding to matching fields of 0.1 to 5
Tesla.  Since TRIM results based on Monte-Carlo simulations give a sharp ion
distribution with an average penetration depth of 56 $\mu$m, our thinned samples
should suffer little ion implantation.

In order to detect  the lower transition of thoriated UBe$_{13}$, we use heat
capacity measurements.  We use an adiabatic method, with a RuO$_2$ thin film
thermometer and a [50:50] AuCr thin film heater.  A fine copper wire provides
the heat link to a dilution refrigerator.  By measuring heat capacity, we
conveniently avoid attaching leads to our tiny samples.  The
unusually large specific heat of the heavy fermions makes the measurements
challenging but not heroic.

\begin{figure}[htb]
\begin{center}
\psfrag{uthbe13}{\large U$_{0.97}$Th$_{0.03}$Be$_{13}$}
\psfrag{tinmk}{\large $T$ (mK)}
\psfrag{c/tinj/molek2}{\large $C/T$ (J/mole K$^2$)}
\includegraphics[scale=1.5]{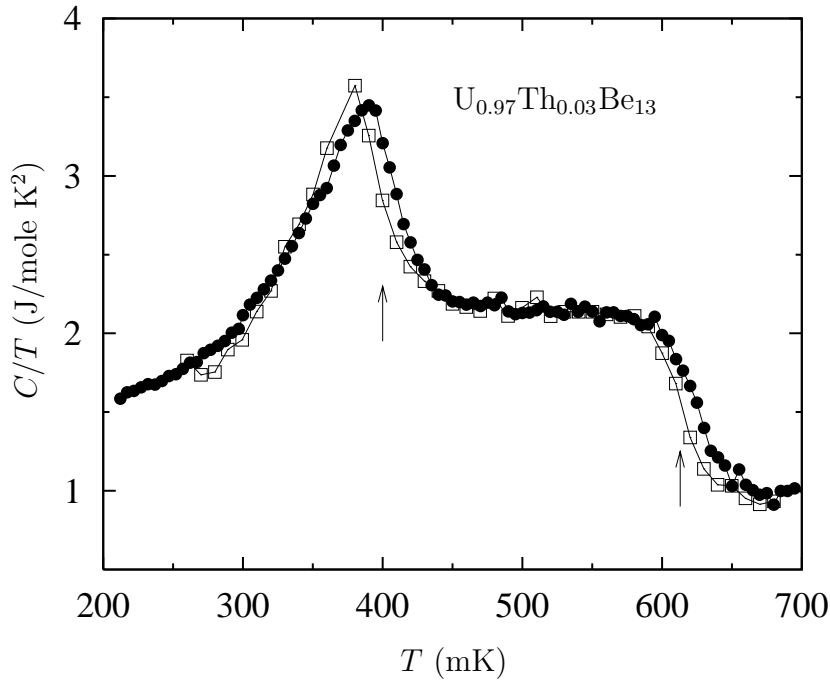}
\caption{Heat capacity $C/T$ for U$_{0.97}$Th$_{0.03}$Be$_{13}$ with ($\Box$)
and without ($\bullet$) heavy-ion irradiation.  The arrows indicate the
two transition temperatures for the irradiated sample.}
\label{f:largenenough}
\end{center}
\end{figure}

Fig. 1 shows $C/T$ for both unirradiated and irradiated
U$_{0.97}$Th$_{0.03}$Be$_{13}$ samples.  For clarity, we show only data from 
the most heavily irradiated sample.  Since we
do not know the precise sample sizes, we use a multiplicative constant to set
the normal state value of $C/T$ to  1 $\mbox{J/mole K}^2$ for each curve. 
We do not adjust the measured heat capacity for any contribution from the
thermometer, heater, or mount; but comparing the curves shown to previous heat
capacity measurements on bulk samples suggests that background effects are less
than 30\% of our signal. In any case, the background should not
affect our determination of the $T_c$ suppression or transition width.

The heat capacity curves of Figure 1 have the same overall shape, including 
the heights of both transitions.  Irradiation has a
similar effect on pure UBe$_{13}$, reducing $T_c$ and slightly broadening
the transition, while retaining the shape of the $C/T$ curve.

Table 1 summarizes the changes in the transitions.  The $T_c$ values correspond
to the midpoints in $C/T$, and are reproducible to within 2 mK for different
cooldowns. The transition widths are taken as the temperature difference
between the 10\% and 90\% points in $C/T$.  We note that the percentage change
in $T_c$ in pure UBe$_{13}$ is nearly identical to that of the lower
transition  in U$_{0.97}$Th$_{0.03}$Be$_{13}$, and somewhat larger than that of
the upper transition.  The two former transitions also undergo a 7 mK  change
in width, while the width of the latter increases by only 3 mK.  These values
suggest that the lower transition in thoriated UBe$_{13}$ is a continuation of
the phase boundary in the pure material.  Work on the pressure-dependence
of the transitions also supported this conclusion\cite{Lambert,Zieve}, 
although recent thermal
expansion measurements did not\cite{Kromer}. The generally similar 
behavior of all three transitions is also evidence that the lower
transition in U$_{0.97}$Th$_{0.03}$Be$_{13}$ is a 
superconducting transition.

Without resistivity data we cannot compare our results to theoretical work
on impurity scattering, but we can examine other experiments.  The $T_c$
reduction from columnar defects in HTS is up to 5.3\% for $10^{15}$
defects/m$^2$ in YBa$_2$Cu$_3$O$_{7-\delta}$ \cite{YBCO} and 2.4\% for
$5\times 10^{14}$ defects/m$^2$ in Bi$_2$Sr$_2$CaCu$_2$O$_8$\cite{BSCCO}.
Since transition temperatures are so different for HTS and HF, we also
consider a recent study in UPt$_3$\cite{Suderow}.  Here point defects
were created by electron irradiation, with most of the electrons passing
through the sample to insure a homogeneous defect density.   If 1\%
of the electrons produced defects, then for the sample sizes given in 
\cite{Suderow}
an electron flux $3\times 10^{22}/{\mbox m}^2$ would yield a final
defect spacing slightly under 20 nm.  The measured $T_c$  reduction for
this flux was 20 mK (3.6\%), slightly more than the reductions in our
highest-density samples, which have 20 nm defect spacing.  Thus our
$T_c$ data is consistent with experiments on other unconventional
superconductors.

\begin{table}
\caption{Effects of irradiation on transitions in (U,Th)Be$_{13}$.}
\label{table:1}
\newcommand{\m}{\hphantom{$-$}}
\newcommand{\cc}[1]{\multicolumn{1}{c}{#1}}
\renewcommand{\tabcolsep}{1pc} 
\renewcommand{\arraystretch}{1.2} 
\begin{tabular}{@{}lcccc}
\hline
   & $T_{c} \mbox{(unirr)}$ & $T_c \mbox{(irr)}$ & $\mbox{width (unirr)}$ & $\mbox{width (irr)}$ \\
\hline
UBe$_{13}$ & 770 & 750 & 35 & 42 \\
U$_{0.97}$Th$_{0.03}$Be$_{13}$ (upper transition) & 625 & 613 & 38 & 41\\
U$_{0.97}$Th$_{0.03}$Be$_{13}$ (lower transition) & 412 & 400 & 37 & 44\\
\hline
\end{tabular}\\[2pt]
All temperatures are given in milliKelvin.
\end{table}

We wish to thank D.L. Cox for helpful discussions and R. Durner
for the TRIM simulation.  This work was supported
by the NSF under DMR-9733898 (UCD) and by DOE grant \#W-31-109-ENG-38
(Argonne).

\end{document}